\documentclass{article}
\usepackage{ijcai25}
\usepackage{times}
\usepackage{soul}
\usepackage{url}
\usepackage[hidelinks]{hyperref}
\usepackage[utf8]{inputenc}
\usepackage[small]{caption}
\usepackage{graphicx}
\usepackage{amsmath}
\usepackage{amsthm}
\usepackage{booktabs}
\usepackage{algorithm}
\usepackage{algorithmicx}
\usepackage[noend]{algpseudocode}
\usepackage[switch]{lineno}

\title{TSS GAZ PTP: Towards Improving Gumbel AlphaZero with Two-stage Self-play for Multi-constrained Electric Vehicle Routing Problems}

\author{
    Hui Wang
    \affiliations
    Affiliation
    \emails
    email@example.com
}

\author{
Hui Wang$^1$
\and
Xufeng Zhang$^1$\and 
Xiaoyu Zhang$^1$\and
Zhenhuan Ding$^1$\and
Chaoxu Mu$^{1,2}$\\
\affiliations
$^1$Anhui University\\
$^2$Pengcheng Laboratory\\
\emails
cxmu@tju.edu.cn
}

\begin{document}

\maketitle
\begin{abstract}

% Since AlphaZero achieved superhuman performance in solving two-player games purely by self-play, a lot of work based on AlphaZero has been investigated, providing a promising approach to tackle combinatorial optimization~(CO) problems. Specifical
Recently, Gumbel AlphaZero~(GAZ) was proposed to solve classic combinatorial optimization problems such as TSP and JSSP by creating a carefully designed competition model~(consisting of a learning player and a competitor player), which leverages the idea of self-play. However, if the competitor is too strong or too weak, the effectiveness of self-play training can be reduced, particularly in complex CO problems. To address this problem, we further propose a two-stage self-play strategy to improve the GAZ method~(named TSS GAZ PTP). In the first stage, the learning player uses the enhanced policy network based on the Gumbel Monte Carlo Tree Search~(MCTS), and the competitor uses the historical best trained policy network~(acts as a greedy player). In the second stage, we employ Gumbel MCTS for both players, which makes the competition fiercer so that both players can continuously learn smarter trajectories. We first investigate the performance of our proposed TSS GAZ PTP method on TSP since it is also used as a test problem by the original GAZ. The results show the superior performance of TSS GAZ PTP. Then we extend TSS GAZ PTP to deal with multi-constrained Electric Vehicle Routing Problems~(EVRP), which is a recently well-known real application research topic and remains challenging as a complex CO problem. Impressively, the experimental results show that the TSS GAZ PTP outperforms the state-of-the-art Deep Reinforcement Learning methods in all types of instances tested and outperforms the optimization solver in tested large-scale instances, indicating the importance and promising of employing more dynamic self-play strategies for complex CO problems.
\end{abstract}


\section{Introduction}

\begin{figure*}[bth!]
\centering
\includegraphics[width=1.0\textwidth]{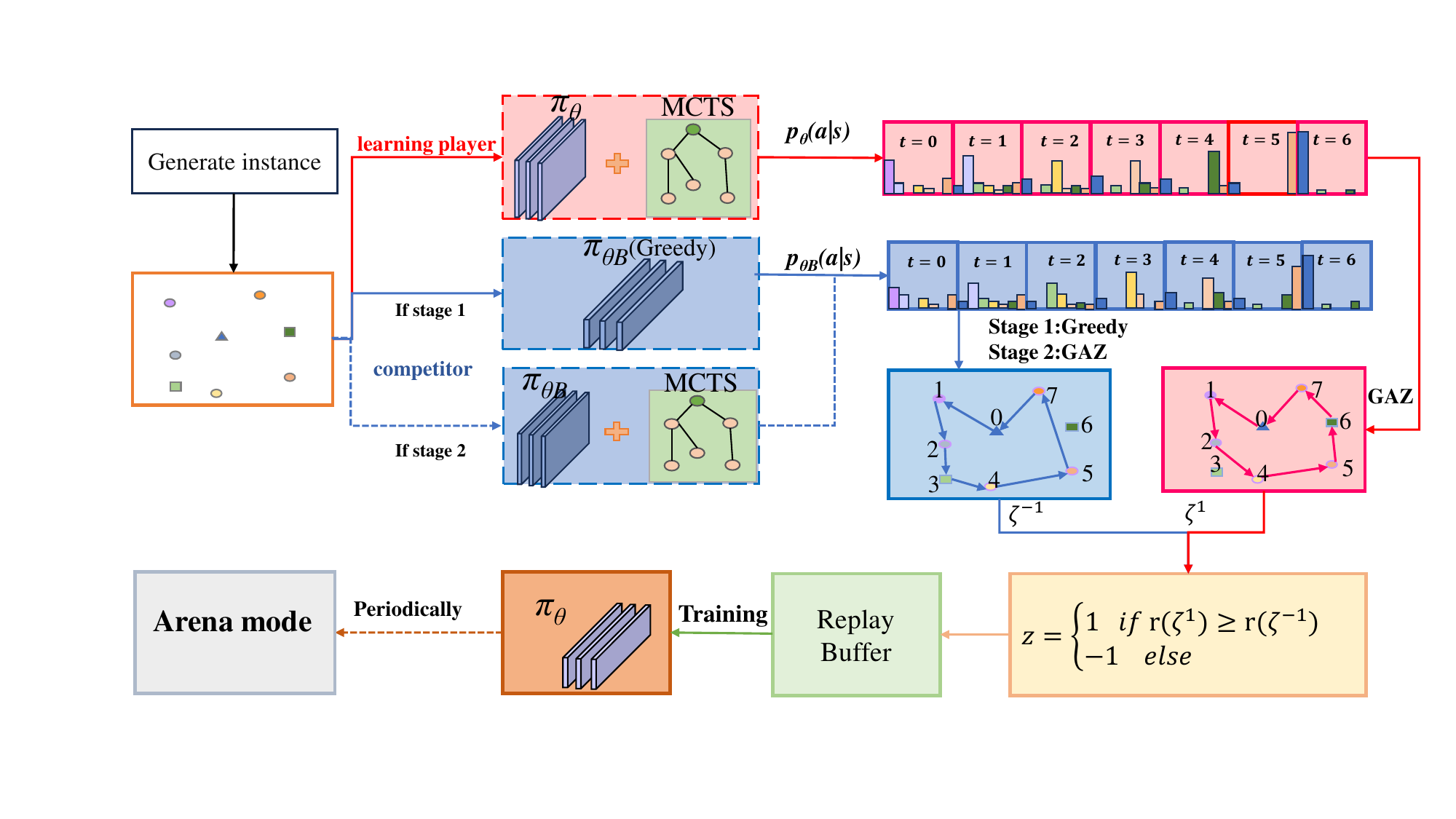} 
\caption{The basic framework of proposed Two-stage Self-play Gumbel AlphaZero~(TSS GAZ PTP). Red part represents the action selection is made by learning player with Gumbel MCTS, blue part represents the action selection is made by competitor player with either greedy strategy in stage 1 or Gumbel MCTS in stage 2.}
\label{figtraining}
\end{figure*}
In recent decades, the development of transportation infrastructure has become vital to the socioeconomic growth of the area, but it has also brought many environmental issues at the same time, especially carbon emissions~\cite{FAN2018673,KUCUKOGLU2021107650}. Path planning, such as the classic Travel Salesman Problem~(TSP) is a typical and simple model to reduce carbon emissions by minimizing the distance traveled. In addiction, Vehicle Routing Problem~(VRP) is usually viewed as a model of logistics service planning. As a classic combinatorial optimization~(CO) problem, it has a lot of variants~\cite{9478307,imran2024smartpathfinder,9458860}. However, most of these variations are built on traditional fuel vehicles. The consumption of fossil fuels remains one of the main contributors to large-scale carbon emissions~\cite{LI2024131041}. However, with the rapid development of electric vehicles and the construction of charging stations, electric vehicles have offered a viable substitute for traditional fossil fuel vehicles~\cite{Conrad2011TheRV}.

Unlike fuel vehicles that have a long range and possess the significant benefit of widespread fuel station deployment, charging electric vehicles is still a problem due to the fact that electric vehicles usually have a range of about 500-700 km~\cite{ZHANG2023118019}. Consequently, it is more challenging to plan the delivery routes with electric vehicles, which leads to a new variant called the Electric Vehicle Routing Problem~(EVRP)~\cite{YANG2025101814}~\cite{MORADI2025125183}. Recently, several methods have been studied to offer optimal solutions for EVRP, including adaptive large neighborhood search (ALNS)~\cite{GOEKE201581,SISTIG2023120915}, variable neighborhood search (VNS)~\cite{9119394}, ant colony optimization~(ACO)~\cite{ZHANG2018404}. These algorithms perform well on small-scale problems and commonly find an approximate optimal solution, but they have to adjust numerous parameters and can easily get stuck in poor local optimal. 

With the advent of deep reinforcement learning~(DRL), the landscape of addressing complex optimization problems has changed significantly to find new avenues for efficient and innovative solutions by training a deep neural network model. Although DRL could provide desirable solutions to both small- and large-scale problems with lower computational costs and outperform traditional techniques. Most of them concentrate on fewer features and less constrained problems, such as the TSP and VRP, which are basically linear programming problems. In comparison, the research on multi-constrained EVRP is more practical and more complex because it is a nonlinear problem that takes into account a wide range of real-world constraints.

As a successful DRL algorithm for large-scale problems such as the game of Go, AlphaZero utilizes Monte Carlo Tree Search~(MCTS) to organize a series of simulations by building a search tree and using statistical information to guide the search process and make the best decisions more likely. However, AlphaZero performs not well if the number of Monte Carlo simulations is too small. To this end,~\cite{Danihelka2022PolicyIB} proposed Gumbel AlphaZero~(GAZ) by using a small number of MCTS simulations, it can achieve better performance and save computational resources. The Gumbel AlphaZero Play-to-Plan~(GAZ PTP) algorithm proposed by~\cite{0547ebb970b949cf8ec4326c525f408a} has also shown its superiority by allowing only one player to use Gumbel MCTS in problems such as TSP and JSSP which have fixed steps~(e.g. the step number is like the city number in TSP). However, GAZ PTP may fail to improve its policy network because one player who employs Gumbel MCTS has such a huge advantage that it is difficult for the other player to find better trajectories, leading to unbalanced competition. To address this problem, we propose a Two-stage self-play strategy to enhance the GAZ algorithm~(see Figure~\ref{figtraining}). We validate its performance in TSP and further extend it to tackle multi-constrained EVRP, which is a variable-step problem~(since the depot and charging stations can be visited multiple times) and consists of the traditional Distance Minimization Electric Vehicle Routing Problem~(DM-EVRP) and the Energy Minimization Electric Vehicle Routing Problem~(EM-EVRP)~\cite{TANG2023121711} with multiple constraints.

Above all, the main contributions of this paper can be summarized as follows. 

\begin{itemize}
\item A Two-stage self-play strategy is proposed to improve the policy network by forcing the learning player to constantly compete with the opponent with similar playing strength.

\item Unlike the original GAZ used for TSP or JSSP with fixed steps, the proposed method is further well extended to solve the variable-step problems like multi-constrained EVRP through self-play. 

\item The proposed TSS GAZ PTP method not only achieved superior performance on problems with fixed planning steps, but also achieved the state-of-the-art performance on both multi-constrained DM-EVRP and EM-EVRP which need variable-steps planning, compared to traditional heuristic solvers and other learning-based approaches, especially on large-scale instances.
\end{itemize}

\section{Related Work}

In recent years, Transformer has shown superior performance in the area of Natural Language Processing~(NLP), particularly machine translation and sequence-to-sequence tasks. A CO problem is essentially a sequence optimization problem, so their solutions can be described as sequential decisions. Therefore, TSP and EVRP, as typical CO problems, are also being studied using DRL methods. For example,~\cite{10608117} proposed a Graph Attention Network~(GAT) based on encoder that can provide high-dimensional node embedding and graph embedding for downstream tasks of EVRP.~\cite{TANG2023121711} formulated an energy consumption model rather than a traditional distance model using the transformer-based DRL method. To address the low efficiency for large-scale EVRP,~\cite{9924526} designed a two-layer model that finds near-optimal solutions based on predefined feasibility conditions and rewards. 

In addition, based on the self-play training strategy, AlphaGo Zero~\cite{Silver2018AGR} achieved superhuman performance on the game of Go.~\cite{Wang2021AdaptiveWM} used MCTS with warm-start enhancements to enhance the quality of the plays produced by self-play. Although AlphaGo Zero was designed for two-player games, many researchers attempted to apply the AlphaZero algorithm to single-player tasks by creating competitive environments. Most of them reconstruct reward mechanisms based on self-competition for different problems~\cite{Bansal2017EmergentCV,laterre2018ranked,9356942}. However, these methods do not use MCTS during training the policy network or value network, they only apply MCTS guided network after training. To address this issue,~\cite{WANG2021104422} improved the policy by employing complete information from the MCTS search tree and learning the trajectory produced by MCTS. To save simulation costs, GAZ was proposed and in order to improve the GAZ-based policy network~\cite{Danihelka2022PolicyIB},~\cite{0547ebb970b949cf8ec4326c525f408a} further developed a new framework. 

Through self-competition, the agent learns to find strong trajectories by planning against potential strategies of its previous self and has shown higher performance on classical CO problems such as TSP and JSSP. But GAZ-based methods have not been investigated yet in multi-constrained EVRP with variable planning steps, which still remains a challenge.

\section{EVRP}
\subsection{Problem Formulation}
\begin{figure}[!t]
\centering
\includegraphics[width=0.9\columnwidth]{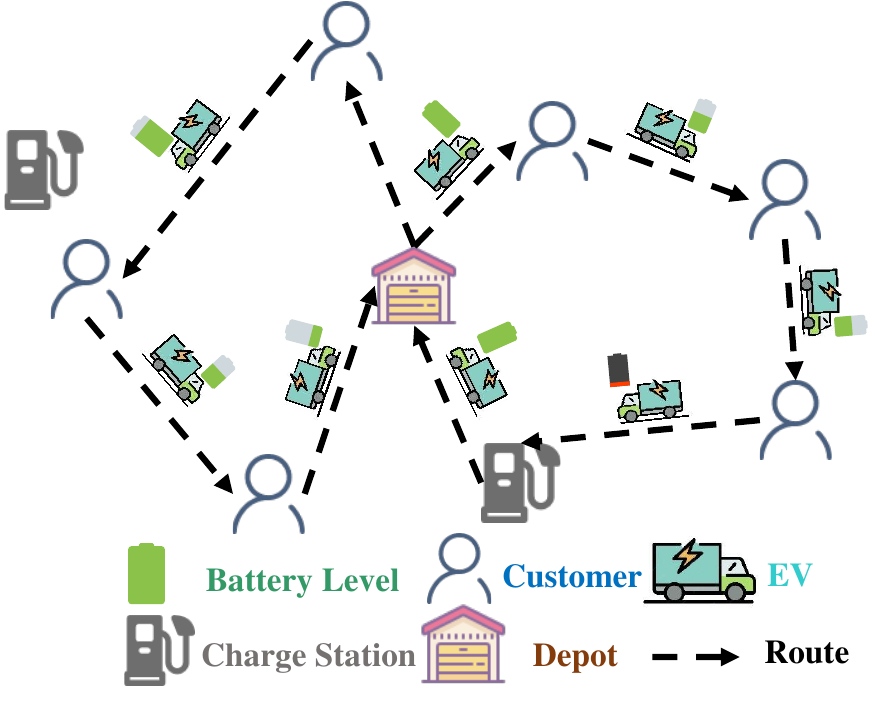} 
\caption{A simple example of the multi-constrained EVRP path planning}
\label{fig1}
\end{figure}

This section introduces the multi-constrained EVRP. Following~\cite{TANG2023121711}, we define it with a directed graph $G =
(V, E)$, $V=C\cup D\cup\hat{F}$, where $C=\{1+s,\ldots,n+s\}$ is a set of $n$ customers, $D=\{0\}$ represents the depot, $\hat{F}=\{1,\ldots,s\}$ is a set of recharge stations, and $E = \{(i,j) \mid i,j \in V\}$ is a set of edges connecting two nodes. Each customer $i$ with demand $c_{i}$ can only be served once. Our goal is to visit all customers and plan completed routes for electric vehicles to minimize the total distance or total energy consumption while satisfying all relevant constraints described in the following equations. As shown in Figure~\ref{fig1}, electric vehicles start from the depot $D$ with a maximum load capacity $L$ and a maximum battery capacity $Q$, and then return to the depot $D$ after serving all customers. During the whole journey, we are supposed to ensure that the capacity of the electric vehicles is not less than 0, and the electric vehicles must visit the recharge stations at the appropriate time to avoid being stranded while driving. We also take into account the maximum serving time of the driver, $T_{max}$.

\begin{flalign}
\label{equation 1}
&\min\quad f(\mathbf{x})=\sum_{i\in V,j\in V,i\neq j}e_{ij}x_{ij}&
\end{flalign}
\begin{flalign}
&\mathrm{s.t.~}&
\nonumber
\end{flalign}
\begin{flalign}
\label{equation 2}
&\sum_{j\in V,i\neq j}x_{ij}=1\quad\forall i\in C&
\end{flalign}
\begin{flalign}
\label{equation 3}
&\sum_{j\in V,i\neq j}x_{ij}\leq1\quad\forall i\in\hat{F}&
\end{flalign}
\begin{flalign}
\label{equation 4}
&\sum_{j\in V,i\neq j}x_{ij}-\sum_{j\in V,i\neq j}x_{ji}=0\quad\forall i\in C\cup\hat{F}&
\end{flalign}
\begin{flalign}
\label{equation 5}
&\tau_{0}=0&
\end{flalign}
\begin{flalign}
\label{equation 6}
&0\leq\tau_\mathrm{i}\leq T_{max},\forall i\in C\cup D\cup\hat{F}&
\end{flalign}
\begin{flalign} 
\label{equation 7}
&\tau_i+\left(g_i+t_{ij}\right)x_{ij}-T_{max}\left(1-x_{ij}\right)\leq\tau_j,\forall i,j\in V,i\neq j&
\end{flalign}
\begin{flalign}
\label{equation 8}
&l_j\leq l_i-c_ix_{ij}+L\left(1-x_{ij}\right)\quad\forall i,j\in V,i\neq j&
\end{flalign}
\begin{flalign}
\label{equation 9}
&0\leq l_i\leq L\quad\forall i\in C\cup D\cup\hat{F}&
\end{flalign}
\begin{flalign}
\label{equation 10}
&u_{0}=L&
\end{flalign}
\begin{flalign}
\label{equation 11}
&e_j\leq e_i-E_{ij}+Q\big(1-x_{ij}\big)\quad\forall i\in C\quad\forall j\in V,i\neq j&
\end{flalign}
\begin{flalign}
\label{equation 12}
&e_i=Q\quad\forall i\in D\cup\hat{F}&
\end{flalign}
\begin{flalign}
\label{equation 13}
&0\leq e_i\leq Q\quad\forall i\in C\cup D\cup\hat{F}&
\end{flalign}
\begin{flalign}
\label{equation 14}
&x_{ij}\in\{0, 1\}\quad\forall i,j\in V,i\neq j&
\end{flalign}
The objective function of the multi-constrained EM-EVRP is depicted in Equation~(\ref{equation 1}). Constraints~(\ref{equation 2}) and~(\ref{equation 3}) stipulate that each customer should only be served once, and the charging stations can be visited multiple times separately. Constraint~(\ref{equation 4}) indicates the route's continuity, while Constraints~(\ref{equation 5}) and~(\ref{equation 6}) ensure that the driver's serving time is no longer than $T_{max}$ and recount after returning to the depot. Constraint~(\ref{equation 7}) states that the driver's serving time is constantly updated when arriving at a site. Constraint~(\ref{equation 8}) tracks electric vehicle cargo, while Constraints~(\ref{equation 9}) and (\ref{equation 10}) ensure that electric vehicles leave the depot fully loaded and have a cargo capacity of $L$. Constraint~(\ref{equation 11}) states that the battery capacity of electric vehicles is constantly updated when they arrive at a site, while Constraint~(\ref{equation 12}) and~(\ref{equation 13}) ensure that the remaining battery capacity is not greater than $Q$ and is fully charged when arriving at the depot or recharge stations. Constraint~(\ref{equation 14}) defines the decision variables. The parameters used in these equations and their corresponding explanation are described in detail in Table~\ref{table1}, see the Appendix. 

The multi-constrained DM-EVRP model shares the same constraints~(equation~\ref{equation 2} to \ref{equation 14}), but the objective function for multi-constrained DM-EVRP is defined as~(\ref{equation 15}).

In addition, we model the multi-constrained EVRP as a two-player game, then the Markov decision
process~(MDP) can be defined as a tuple $(\mathcal{S},\mathcal{A},\mathcal{R},\mathcal{P}) $, consisting of state $\mathcal{S}$, action $\mathcal{A}$, reward $\mathcal{R}$ and state transition $\mathcal{P}$, see the detailed definition of each element in Appendix.

\begin{flalign}
  \label{equation 15}
&\min\quad f(\mathbf{x})=\sum_{i\in V,j\in V,i\neq j}d_{ij}x_{ij}&
\end{flalign}
\subsection{Energy Consumption}

We calculate the energy consumption of the electric vehicles between node i and node j as follows:
\begin{equation}
\label{equation 16}
    P_{ij}=\left(m_{ij}\left(a+g sin\left(\alpha_{ij}\right)+C_{r} cos\left(\alpha_{ij}\right)\right)+S\right)\nu_{ij}
\end{equation}
\begin{equation}
\label{equation 17}
    S=0.5\cdot C_d\cdot\rho\cdot A\cdot{\nu_{ij}}^2
\end{equation}
where $m_{ij}$ represents the capacity of the electric vehicles between the node $i$ and node $j$, $a$ is the acceleration of the vehicle and is set as 0 since we assume the speed is constant in our experiments. Gravity is shown by $g$, air density is indicated by $\rho$, resistance coefficient is indicated by $C{r}$, aerodynamic drag coefficient is indicated by $C_{d}$, and slope between nodes $i$ and $j$ is indicated by $\alpha_{ij}$. After getting the mechanical power $P_{ij}$, the energy consumption is calculated as follows:
\begin{equation}
\label{equation 18}
\left.e_{ij}=\left\{\begin{array}{c}\phi^d\cdot\varphi^d\cdot P_{ij}\cdot t_{ij} \quad\mathrm{P}_{ij}\big(m_{ij}\big)\geq0 kw\\\phi^r\cdot\varphi^r\cdot P_{ij}\cdot t_{ij} \quad\mathrm{P}_{ij}\big(m_{ij}\big)<0 kw\end{array}\right.\right.
\end{equation}
Since the effect of slope is a factor, the formula for energy consumption is divided into two cases: electric vehicles need more energy when traveling uphill and are allowed to charge for power recovery when traveling downhill. While $t_{ij}$ represents the travel time between node $i$ and node $j$, $\phi^d$ and $\phi^r$ denote the charging and discharging efficiency of the battery, respectively.

\section{Methodology}
\subsection{Two-stage Self-play}
\begin{figure}[!t]
\centering
\includegraphics[width=1.0\columnwidth]{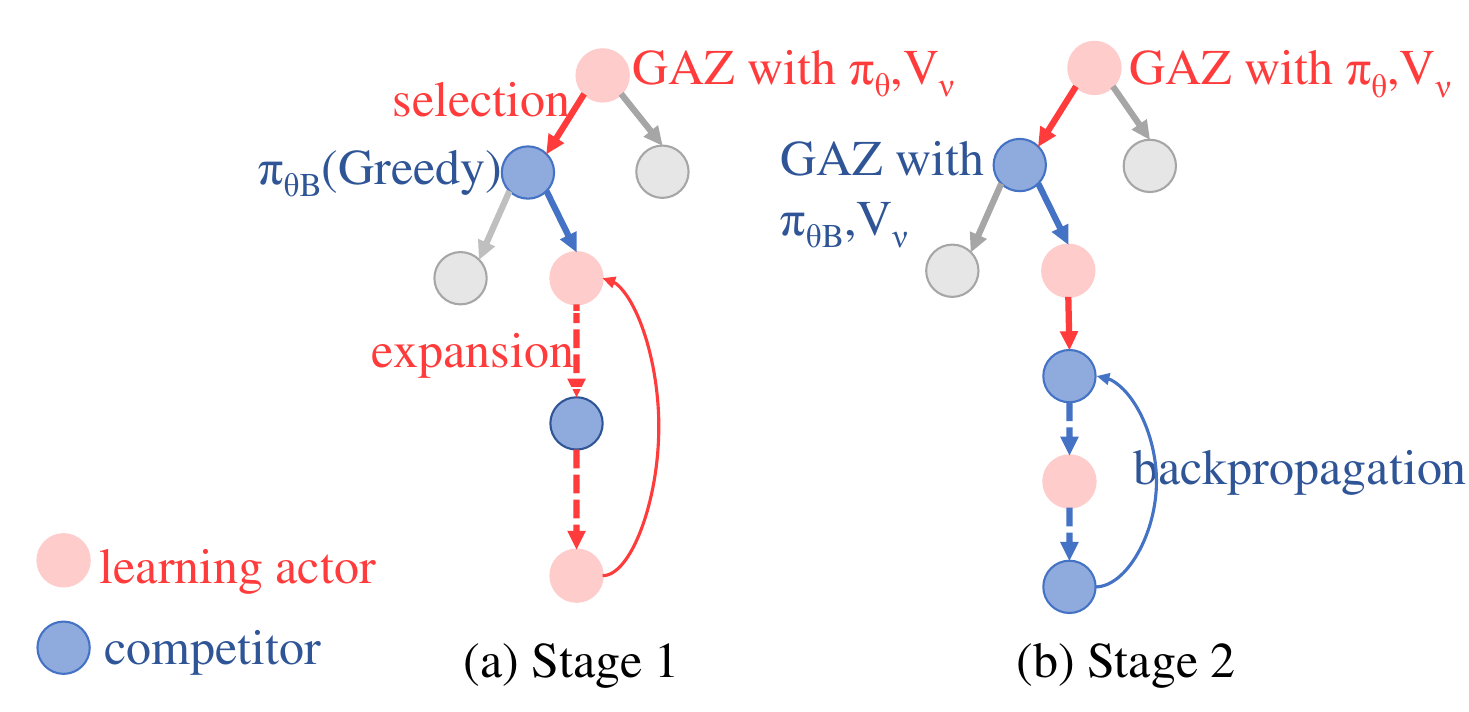} 
\caption{The comparison of Gumbel MCTS for Stage 1 and 2}
\label{fig4}
\end{figure}

The pseudo code of the original GAZ PTP method can be found in the Appendix, Algorithm~\ref{alg:training2}. This part presents our new GAZ method with Two-stage self-play strategy. As shown in Figure~\ref{fig4}, on stage $1$, the learning player uses Gumbel MCTS to choose actions, while the competitor chooses the action from the best historical policy network. In this stage, only the learning player can take the state of the competitor into consideration when in the expansion phase and update the node information through backpropagation. After a period of training episodes, the learning player is unable to find a better trajectory because it has a great advantage to employ Gumbel MCTS in complex tasks, leading to unbalanced competition. Therefore, we introduce the second stage. Both players use Gumbel MCTS, which ensures that they try to find a better trajectory during the competition. It also increases the depth of MCTS because both players can take the other player's state into account and update the node information through backpropagation.

\subsection{Algorithm}
The algorithm~\ref{alg:training} illustrates our proposed training framework.
In each episode, we divide the two players into the learning player and the competitor, while we also divide the training into two stages. In the first stage, the learning player chooses the action according to the policy network $\pi_\theta$ based on GAZ MCTS~\cite{Danihelka2022PolicyIB}, while the competitor chooses the action greedily from the policy network $\pi_{\theta^{B}}$, which is the best historical policy network of the learning player. The $\pi_{\theta^{B}}$ only updates periodically in the arena mode when the outcome of greedily rolling out $\pi_\theta$ is better than $\pi_{\theta^{B}}$. Our aim is to train the model to converge quickly and save computational resources simultaneously due to the competitor does not employ Monte Carlo simulations and Gumbel MCTS only needs fewer simulations. After $n$ episodes of training, the learning player performs much better than the competitor.

Therefore, to ensure that the learning player learns stronger trajectories, in the second stage, each player uses the Gumbel MCTS based policy network $\pi_\theta$ and $\pi_{\theta^{B}}$ to select actions. Our proposed Two-stage Self-play strategy can make the learning player constantly compete with an opponent with similar playing strength so that the learning player can learn smarter trajectories from games. In each episode, there is a probability $P$ of performing self-play that helps the learning player learn the trajectories of its own policy and also maintains the diversity of the training data. A general visualization of the framework of this proposed algorithm is described in Figure~\ref{figtraining}.
\begin{algorithm}[!t]
    %\KwData{}
    \caption{Gumbel AlphaZero Play-to-Plan with Two-stage Self-Play}
   \textbf{Input}: {$\rho_0$: initial state distribution; $\mathcal J_{\text{arena}}$: set of initial states sampled from $\rho_0$} \\
    \textbf{Input}:{$0 \leq P < 1$: self-play parameter}\\
    Init policy replay buffer $ B_\pi = \{\} $ and value replay buffer $ B_V = \{\}$ \;\\
      Init parameters $\theta$, $\nu$ for policy net $\pi_\theta \colon \mathcal S \to \Delta\mathcal A$ and value net $V_\nu \colon \mathcal S \times \mathcal S \to [-1, 1]$ \;\\
    Init 'best' parameters $\theta^B \gets \theta$ \;\\
    Init stage $g= 1$
    \begin{algorithmic}[1]
    \For{\text{episode} $= 1, \ldots, N$}
    \State Assign learning player: $l \gets \text{random}(\{1, -1\})$ \;
    \If {$P \leq \text{random}$ (0,1)$ $}
        \State  $\mu \gets \pi_{\theta} $
    \Else
        \State  $\mu \gets \pi_{\theta^{B}} $
    \EndIf
        \For{\text{step} $t= 0, \ldots, T-1$}
            \For{\text{player} $p = -1, 1,$}
                \If {\text{player} $p\neq l$ \text{and} \text{stage} $g= 1$}
                    \State{\text{Player choose action $a_t^p$ according to policy }$\mu$} 
                    \State{and update state $s_{t+1}^p$}
                \Else
                    \State{Performing policy improvement ${\mathcal{I}}\pi(s_t^p)$} 
                    \State{and using $V_\nu$ and $\pi_\theta$ based on MCTS to} 
                    \State{choose action $a_t^p$ and update state $s_{t+1}^p$}
                    \State{\text{Store}$(s_t^p,{\mathcal{I}}\pi(s_t^p))$ in policy replay buffer} 
                    \State{$\mathcal M_V$}
                \EndIf
           \EndFor
        \EndFor
        \State{Trajectories $\zeta^p\leftarrow(s_0^p,a_0^p,\ldots,s_{T-1}^p,a_{T-1}^p,s_T^p)$ for player $p\in\{1,-1\}$}
        \If {$r(\zeta^1)\geq r(\zeta^{-1})$}
        \State {\text{Game outcome  $z = 1$}}
        \Else
        \State {\text{Game outcome  $z = -1$}}
        \EndIf
        \State{Store tuples $(s_t^1,s_t^{-1},z)$ and $(s_t^{-1},s_{t+1}^1,-z)$ in value replay buffer $\mathcal M_V$}
        \If{$\sum_{s_0\in\mathcal{J}_{\mathrm{arena}}}\left(r(\zeta_{0,\pi_\theta}^{\mathrm{greedy}})-r(\zeta_{0,\pi_{\theta^B}}^{\mathrm{greedy}})\right)>0$}
        \State{\text{Update $\theta^B\leftarrow\theta $ }}
        \EndIf
        \If{\text{episode} $E \geq n$}
            \State{\text{Set stage $g=2 $ }}
        \EndIf
     \EndFor
    \end{algorithmic}
    {\label{alg:training}}
\end{algorithm}

\subsection{Network Architecture}
\begin{figure}[t]
\centering
\includegraphics[width=1.0\columnwidth]{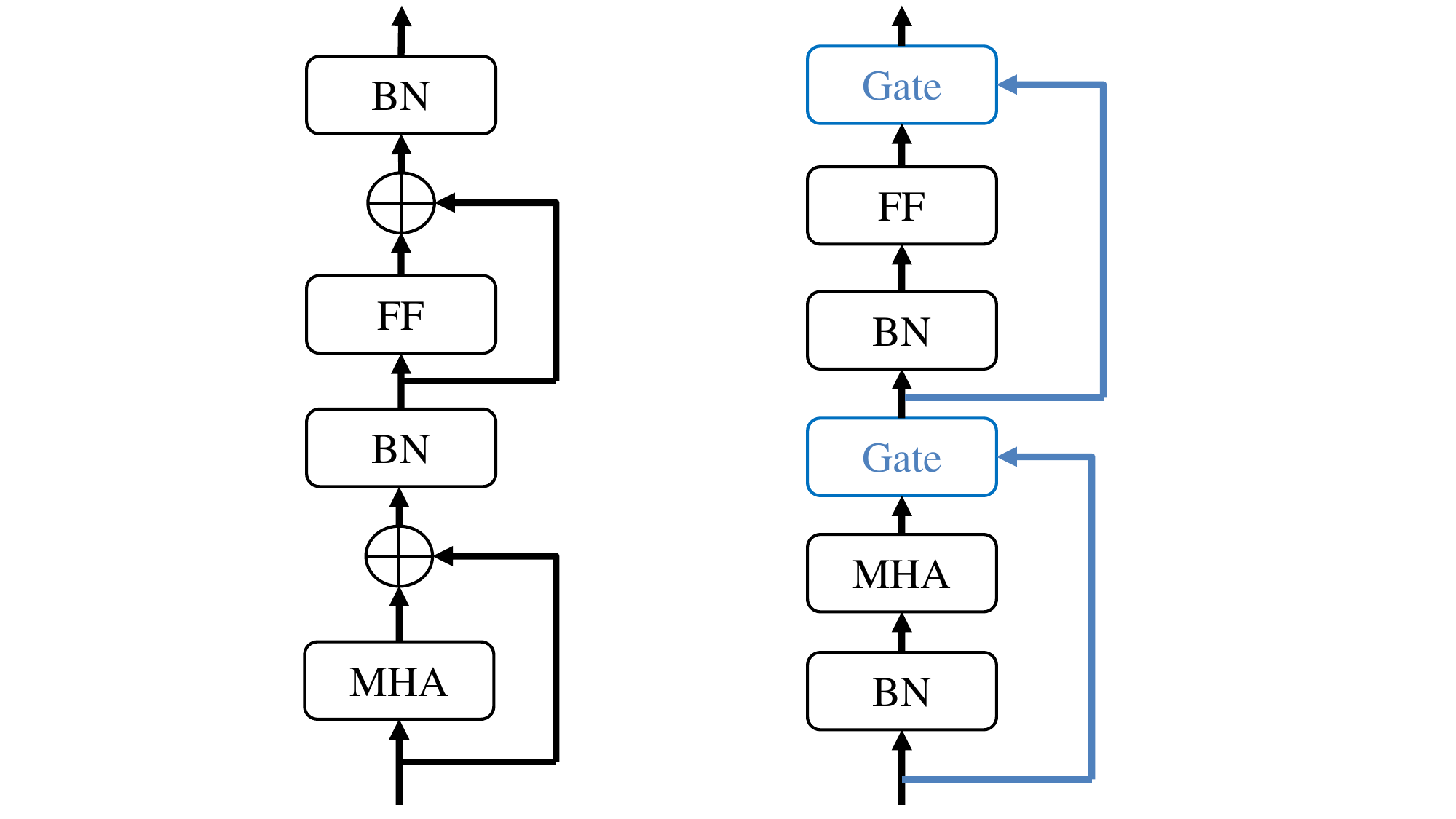} 
\caption{Vanilla Transformer block~(left) and our Transformer block~(right) that adds gate aggregation}
\label{figtransblock}
\end{figure}

The policy and value network are based on the Transformer architecture and the Transformer block is a little different from the Vanilla Transformer block, as shown in Figure~\ref{figtransblock}. We employ batch normalization~(BN) before the Multi-Head Attention~(MHA) and add gate aggregation~\cite{9478307} after MHA and feedforward network~(FFN) to replace the additive aggregation. For the policy network, we use a pointing mechanism based on state attention to compute the probability of each legal action, which is similar to the way in~\cite{Kool2018AttentionLT}.

\section{Experiments}
\subsection{Validation on TSP}
In order to compare the performance of our proposed TSS GAZ PTP and the original GAZ method, we first performed experiments on TSP instances with 20, 50 and 100 nodes, which are also tested by~\cite{Kool2018AttentionLT} and~\cite{0547ebb970b949cf8ec4326c525f408a}. The coordinates for each instance are sampled from $[0,1]^2$. For problems of different scales, we use the same hyper-parameter settings based on GAZ PTP~\cite{0547ebb970b949cf8ec4326c525f408a}. As shown in the Tabel~\ref{Table 5}, the experimental results demonstrate that our method achieves the best performance compared to other historically best learning-based methods on TSP problems.

\begin{table}[!h]
\centering
\caption{TSS GAZ PTP VS Baselines for TSP, each size of the test set consists of \textbf{10000 instances}.}
\renewcommand{\arraystretch}{1.55}
\resizebox{.48\textwidth}{!}{%
   \begin{tabular}{c l|cc|cc|cc}
   \toprule
   & Method        & Obj. & Gap       & Obj. & Gap        & Obj. & Gap \\
   \hline
   & \multicolumn{1}{c}{}  & \multicolumn{2}{c}{$n$ = 20}  & \multicolumn{2}{c}{$n$ = 50} & \multicolumn{2}{c}{$n$ = 100} \\
   \cline{1-8}
   & Optimal Solver~(Concorde)             & 3.84 &   0.00\%        & 5.70        & 0.00\%    & 7.76        & 0.00\% \\
   \cline{1-8}
   & Kool(Attention)   & 3.85 &   0.34\%        & 5.80        & 1.76\%    & 8.12        & 4.53\% \\
   %\cline{1-8}
   & GAZ PTP      & 3.84 & 0.17\% & 5.78 & 1.55 \% & 8.01 & 3.16\% \\
   & TSS GAZ PTP~(ours)    & \textbf{3.84} & \textbf{0.15}\% & \textbf{5.76} & \textbf{1.23}\% & \textbf{7.97} & \textbf{2.71}\% \\
\bottomrule
\end{tabular}%
}
\label{Table 5}
\end{table}

\subsection{Extension to EVRP}
Since the advantage of our TSS GAZ PTP method on TSP has been validated, our aim is to extend the method to variable-step problems. Therefore, we further conducted experiments on multi-constrained EVRP instances with n = 10, 20 and 50 nodes, each category consists of 512 different instances, which is the same as~\cite{TANG2023121711}. For example, instances with 10 and 20 customers have four recharge stations, while instances with 50 customers have eight. Both customer sites and recharge stations are uniformly distributed in the $[0,100]^2$ kilometer area, and the depot is randomly distributed in the $[25,75]^2$ kilometer area. The demand of each customer is uniformly distributed between ${0.25,0.5,0.75,1}$. The specific parameters and their description of the vehicle are shown in Table~\ref{Table 2}. For different scales of problems, we use the same hyper parameter settings based on GAZ PTP~\cite{0547ebb970b949cf8ec4326c525f408a}. The node embedding dimension is 128 and the batch size is 256. We employ the Adam
optimizer with a constant learning rate of $10^{-4}$.
\begin{table}[!t]
\centering
\caption{The parameter values and description for  multi-constrained EVRP setup} 
\begin{tabular}{lcl}
\hline
Parameter    &  Description &Value\\
\hline
%  vehicle attributes
$L$ & Capacity of the vehicle & 4000kg\\
$m_c$ & unladen load & 4100kg\\ 
$Q$ & Battery capacity of the vehicle & 80kwh \\
$A$ & Frontal surface area & $3.912m^2$  \\
$\rho$ & Atmospheric density & $1.2kg/m^3$\\
$g$ & Gravitational constant &  $9.81m/s^2$  \\
$c_r$ & Resistance coefficient &0.01\\
$c_d$ &  Aerodynamic drag coefficient &0.7\\
$\phi^d$ & Propulsion efficiency&1.18\\
$\phi^r$ & Regenerative braking efficiency&0.85\\
$\varphi^d$ &  Charging efficiency &1.11\\
$\varphi^r$ & Discharging efficiency&0.93\\
\hline
\label{Table 2}
\end{tabular}
\end{table}
\begin{figure}[b!]
\centering
\includegraphics[width=1.0\columnwidth]{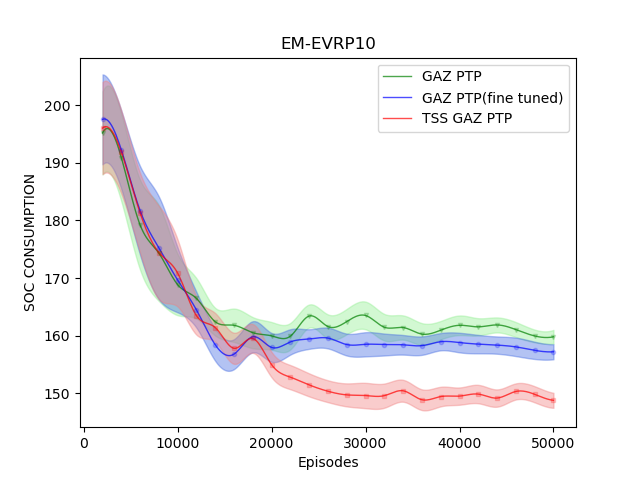} 
\caption{Results of comparison experiments among GAZ PTP, GAZ PTP~(fine tuned) and TSS GAZ PTP on category C10-S4 for EVRP. Our proposed method achieves the lowest SOC consumption. And SOC consumption drastically decreased after 20K episodes, indicating that our proposed method can get rid of the local optimal and achieve better performance.}
\label{fig5}
\end{figure}
\subsection{Baselines}
We compare the proposed TSS GAZ PTP
framework with the following methods.
\begin{itemize}
\item Gurobi: A commercial optimization solver.
\item ACO: An improved ant colony algorithm based meta-heuristics to solve EVRP.~\cite{ZHANG2018404}
\item ALNS: Adaptive large neighborhood search algorithm, which is enhanced by a local search for intensification to solve EVRP.~\cite{GOEKE201581}
\item AM: A Reinforcement Learning method based on attention mechanism.~\cite{Kool2018AttentionLT}
\item DRL: A DRL method with Transformer specifically for EVRP.~\cite{TANG2023121711}
\item GAZ PTP: A Reinforcement Learning method based on self-competition.~\cite{0547ebb970b949cf8ec4326c525f408a}
\item GAZ PTP~(fine tuned): The framework is the same as GAZ PTP, but we have fine-tuned parameters for multi-constrained EVRP.
\end{itemize}

\subsection{Results on EVRP}
Our techniques are implemented on two NVIDIA GeForce RTX 4090 GPUs and an Intel i9-14900K CPU running at 6.00 GHz. With a total of 50k training episodes, after the light-weight test, we divided our training into two stages: 20k and 30k, with a simulation budget of 100 since~\cite{TANG2023121711} and~\cite{Kool2018AttentionLT} train on 128M trajectories. The following are the training durations for certain problems: 15 hours for C10-S4, 40 hours for C20-S4, and 120 hours for C50-S8. 

To present the training curve, we select the category C10-S4 EM-EVRP as an example and present the results in Figure~\ref{fig5}. We see that along with training, total energy consumption gradually decreases. For our TSS GAZ PTP method, SOC consumption drastically decreased after 20K episodes, demonstrating the efficacy of the second training phase. The training curve eventually converges, indicating that, despite sporadic oscillations during the training phase, our approach has learned the best consistent policy.

%\subsection{Analysis}
\begin{figure}[b!]
\centering
\includegraphics[width=1.0\columnwidth]{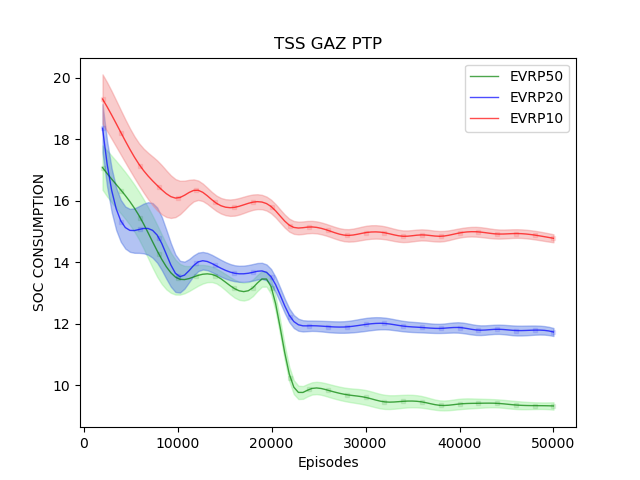} 
\caption{
TSS GAZ PTP training curves on instances of category C10-S4, category C20-S4, and category C50-S8. For comparison, we divide the SOC consumption of EVRP for each category by the number of customers to obtain the average battery consumption per customer. The results indicates the significant improvement of the proposed method especially for larger problems.}
\label{fig6}
\end{figure}
% Moreover, Figure~\ref{fig5} shows the experimental results for multi-constrained EVRP in comparing GAZ PTP, our fine tuned GAZ PTP, and our proposed method TSS GAZ PTP on category C10-S4. During the training process, we found that the original model is not efficient enough to learn the correlation between features, because multi-constrained EVRP is more complicated and challenging and has much more constraints and states compared to TSP. Thus, we adjusted the self-play parameter from 0.2 up to 0.5 and increased Transformer blocks from 5 to 6. We compare the learning curves of GAZ PTP and GAZ PTP~(fine tuned) to investigate if modifying these parameters helps to enhance the performance of the model. As depicted in figure~\ref{fig6}, we can see that GAZ PTP~(fine tuned) achieves
% better learning compared to the original GAZ PTP. However, it
% is still not as good as our proposed TSS GAZ PTP.

\begin{table*}[!t]
\caption{TSS GAZ PTP VS Baselines for Multi-constrained EM-EVRP and DM-EVRP, each category of the test set has \textbf{512 instances}.}
\centering
{
\begin{tabular}{c|c c|c c|c c} 
\hline
\toprule
& 
\multicolumn{2}{c|}{EM-EVRP10} &
\multicolumn{2}{c|}{EM-EVRP20} & 
\multicolumn{2}{c}{EM-EVRP50} \\ 
Method & Obj.~(kwh)   & Gap     & Obj.~(kwh)  & Gap    & Obj.~(kwh)   & Gap        \\  \hline
Gurobi  & 145.25 & 0\%  & 225.52 & 0\%		& -	& -  	 \\
 \hline
ALNS & 151.15 & 4.04\%   & 241.54 & 7.19\%	& 480.64	& 3.61\%  	 \\
ACO &  151.98 & 4.61\%    & 240.15 & 6.49\%	 & 476.46	& 2.71\%  	 \\  \hline
AM~(Greedy)&  152.87 &5.22\%   & 247.65 & 9.81\%		& 486.54	& 4.88\%  	 \\
AM~(Sample1280)  &149.62 & 2.99\%   & 238.63 & 5.81\%		& 471.54	& 1.65\%  	 \\
AM~(Sample12800)  &149.28 & 2.75\%   & 237.56 & 5.34\%		& 468.46	& 0.99\% 	 \\
DRL~(Greedy)	&  151.35 & 4.18\%   & 244.54 & 8.43\%		& 482.27	& 3.96\%  \\
DRL~(Sample1280)  &  148.77 & 2.40\%    & 237.43 & 5.28\%		& 466.70	&0.61\% \\  
DRL~(Sample12800)  &  148.43 & 2.17\%    & 236.33 & 4.79\%		& 464.69	&0.18\%\\  
GAZ PTP  &  160.14 & 10.25\%    & 252.27 & 11.86\%		& 570.18	& 22.92\%  \\  
GAZ PTP~(fine tuned)  &  155.23 & 6.84\%    & 245.85 & 9.01\%		 & 550.75	& 18.73\%  \\  
TSS GAZ PTP  &  \textbf{147.79} & \textbf{1.72}\%    & \textbf{234.68} & \textbf{4.06}\%		 & \textbf{463.85}	& \textbf{0.00\%} \\   \hline
% \bottomrule
% \end{tabular}}
% \label{Table 3}
% \end{table*}
% \begin{table*}
% \caption{TSS GAZ PTP VS Baselines for multi-constrained DM-EVRP.}
% \centering
% % \setlength{\abovecaptionskip}{0.2cm}
% % \setlength{\belowcaptionskip}{-0.3cm}
% {
% \begin{tabular}{c|c c|c c|c c} 
% \hline
% \toprule
& 
\multicolumn{2}{c|}{DM-EVRP10} &
\multicolumn{2}{c|}{DM-EVRP20} & 
\multicolumn{2}{c}{DM-EVRP50} \\ 
 & Obj.~(km)   & Gap     & Obj.~(km)   & Gap    & Obj.~(km)   & Gap \\ \hline
Gurobi  & 346.59 & 0\%  & 542.55 & 0\%		& -	& -\\
 \hline
ALNS & 355.01 & 2.43\%   & 572.21 & 5.47\%	& 1147.31	& 4.26\%   \\
ACO &  353.85 & 2.09\%    & 568.15 & 4.72\%	 & 1140.74	& 3.66\%   \\ \hline
AM~(Greedy)&  361.76 &4.37\%   & 584.14 & 7.67\%	& 1153.05	& 4.78\% \\
AM~(Sample1280)  &357.23 & 3.07\%   & 571.45 & 5.33\%		& 1118.94	& 1.68\% \\
AM~(Sample12800)  &355.01 & 2.43\%   & 568.34 & 4.75\%		& 1112.04	& 1.05\%  \\
DRL~(Greedy)	&  357.24 & 3.07\%   & 577.38 & 6.42\%		& 1139.56	& 3.55\%  \\
DRL~(Sample1280)  &  352.72 & 1.77\%    & 565.17 & 4.17\%		& 1108.74	&0.75\%\\  
DRL~(Sample12800)  &  352.35 & 1.66\%    & 562.16 & 3.61\%		& 1104.03	&0.32\% \\  
GAZ PTP  &  380.33 & 9.83\%    & 599.84 & 10.56\%		& 1328.14	& 20.69\% \\  
GAZ PTP~(fine tuned)  &  366.07 & 5.62\%    & 582.05 & 7.28\%		 & 1283.49	&16.63\%  \\  
TSS GAZ PTP  &  \textbf{351.84} & \textbf{1.51}\%    & \textbf{560.49} & \textbf{3.30}\%		 &  \textbf{1100.47}	&\textbf{0.00\%} \\   \hline
\bottomrule
\end{tabular}}
\label{Table 4}
\end{table*}
In addition, we conducted convergence analysis experiments on the complete learning process of the TST GAZ PTP, visualizing the reduction of SOC consumption as the number of training iterations increases. In Figure~\ref{fig6}, the x-axis represents the number of iterations, and the y-axis represents the average SOC consumption for visiting each customer. The decreasing SOC consumption with increasing training iterations demonstrates the effectiveness of the TST GAZ PTP algorithm on all types of EVRP instances. Importantly, we see that for larger instances, our TSS GAZ PTP achieves more significant improvements, indicating its potential to handle larger size problems.

The complete results of the comparison experiments for instances of different sizes are shown in Tables~\ref{Table 4}. Similar to~\cite{TANG2023121711}, we measure performance by the
total energy consumption $E$ and the normalized energy consumption between the total energy consumption $E$ and the best objective value $E_\mathrm{best}$ across all methods. We see that in category C10-S4 and C20-S4, the Gurobi solver performs better than other methods, but fails to find optimal solutions to large-scale problems. The proposed TSS GAZ PTP performs better than other learning-based models both in multi-constrained EM-EVRP and DM-EVRP on all different size instances. As the number of nodes increases, its advantage becomes increasingly notable. We can see that TSS GAZ PTP significantly outperforms the previous state-of-the-art
methods on all C50-S8 instances, indicating the potential of dealing with large-scale multi-constrained problems.

\section{Conclusion}
In this paper, we proposed TSS GAZ PTP, a novel DRL framework inspired by GAZ, for solving multi-constrained EVRP. Unlike reconstructing network architecture or adding state feature modules for specific tasks, our aim is to design a new training strategy based on self-play that can be applied to more complex and general tasks and enhance the diversity of training data. Our training strategy ensures that the player continues to explore a smarter trajectory based on self-play and provides an efficient way to eliminate local optimal. The experimental results show that our method outperforms both current state-of-the-art DRL methods and heuristic algorithms, especially in large-scale instances. Importantly, the proposed method achieved a significant improvement compared to the original GAZ PTP method on multi-constrained EVRP.

For future work, it is still promising to further explore more efficient self-play strategies for GAZ-type methods and apply these methods to solve more multi-constrained tasks, especially for their large-scale instances. Another direction is to explore alternative ways to reduce the computational cost caused by Monte Carlo simulations in large-scale problems. In addition, combining the proposed approach with the multi-agent methods~\cite{Hao2024MultiagentGM} for more complex tasks requires more investigation.

\bibliographystyle{named}
\bibliography{ijcai25}

\clearpage
\section{Appendix}
\subsection{GAZ PTP}

According to Algorithm~\ref{alg:training2}, we see that the original GAZ PTP algorithm only has one stage of self-play. Throughout the training process, the learning player selects actions using a policy network guided by MCTS planning, while the greedy player greedily selects the action with the highest probability according to the policy network. The greedy player's strategy performs reasonably well in problems like TSP, which only requires considering the simple constraint that the next selected point has not been visited before. However, for problems with multiple constraints like EVRP, the greedy player's greedy strategy is insufficient to execute a near-optimal action. This leads to a situation where the competition between the learning player and the greedy player fails to produce better-learned strategies.
 \begin{algorithm}[!h]
    %\KwData{}
    \caption{GAZ Play-to-Plan (GAZ PTP) Training}
   \textbf{Input}: {$\rho_0$: initial state distribution; $\mathcal J_{\text{arena}}$: set of initial states sampled from $\rho_0$} \\
    \textbf{Input}:{$0 \leq \gamma < 1$: self-play parameter}\\
    Init policy replay buffer $\mathcal M_\pi \gets \emptyset$ and value replay buffer $\mathcal M_V \gets \emptyset$ \;\\
      Init parameters $\theta$, $\nu$ for policy net $\pi_\theta \colon \mathcal S \to \Delta\mathcal A$ and value net $V_\nu \colon \mathcal S \times \mathcal S \to [-1, 1]$ \;\\
    Init 'best' parameters $\theta^B \gets \theta$ 
    \begin{algorithmic}[1]
    \For{\text{episode} }
    \State Sample initial state $s_0 \sim \rho_0$ and set $s_0^p \gets s_0$ for $p = 1, -1$ \;\\
    Assign learning actor to player position: $l \gets  \text{random}(\{1, -1\})$ \;\\
         Set greedy actor's policy $\mu\leftarrow\begin{cases}\pi_{\theta} \\\pi_{\theta^{B}}\end{cases}$\\
        \For{\text{step} $t= 0, \ldots, T-1$}
            \For{\text{player} $p = -1, 1,$}
                \If {\text{player} $p\neq l$ }
                    \State{\text{Take greedy action $a_t^p$ according to policy } 
                    \State{ $\mu(s_t^p)$and receive new state $s_{t+1}^p$}}
                \Else
                    \State{Perform policy improvement $\mathcal I$ with MCTS using $V_\nu(\cdot, \cdot)$ and $\pi_\theta(\cdot)$ for player $p$ \;where in tree, player $-p$ samples (resp. chooses greedily) actions from $\mu$ \;} 
                    \State{Receive improved policy $\mathcal I\pi(s_t^p)$, action $a_t^p$ and new state $s_{t+1}^p$\;
                     Store $(s_t^p, \mathcal I\pi(s_t^p))$ in $\mathcal M_\pi$\;} 
                \EndIf
           \EndFor
        \EndFor
        \State{Have trajectories $\zeta^p \gets (s_0^p, a_0^p, \dots, s_{T-1}^p, a_{T-1}^p, s_T^p)$ for players $p \in \{1, -1\}$\;}
        \State{
         $z \gets \begin{cases}
            1 & \text{if } r(\zeta^1) \geq r(\zeta^{-1}), \\
                           -1 & \text{else}
                            \end{cases}$}
        \State{Store tuples $(s_t^1, s_t^{-1}, z)$ and $(s_t^{-1}, s_{t+1}^1, -z)$ in $\mathcal M_V$ for all timesteps $t$ \;
         \;}
        \State{Periodically update $\theta^B \gets \theta$ if}
        \State{$\sum_{s_0 \in \mathcal J_{\text{arena}}} \left(r(\zeta_{0, \pi_\theta}^{\text{greedy}}) - r(\zeta_{0, \pi_{\theta^B}}^{\text{greedy}})\right) > 0$ \;}
     \EndFor
    \end{algorithmic}
    {\label{alg:training2}}
\end{algorithm}
\subsection{EVRP Parameters}

\begin{table}[!h]
\centering
\caption{The parameter and description of the multi-constrained EVRP} 
\begin{tabular}{p{0.1\linewidth}p{0.83\linewidth}}
\hline
Parameter    & \qquad Description\\
\hline
%  vehicle attributes
$L$ & Capacity of the vehicle \\
$T_{max}$ & the driver's maximum serving time \\
$Q$ & Battery capacity of the vehicle \\
$c_i$ & Demand of the customer $i$  \\
$l_i$ & Remaining capacity while reaching node $i$  \\
$\tau_i$ & Travel time while reaching node $i$   \\
$g_i$ & Serving time while reaching node $i$  \\
$e_i$ & Remaining battery capacity while reaching node $i$  \\
$d_{ij}$ & Distance between node $i$ and node $j$  \\
$t_{ij}$ & Travel time of the node $i$ from the node $j$\\
$E_{ij}$ & Energy consumption from the node $i$ to $j$ \\
$u_{ij}$ & Cargo load from the node $i$ to the node $j$\\
$v_{ij}$ & Average speed from the node $i$ to the node $j$ \\
\hline
\end{tabular}
\label{table1}
\end{table}
%\subsubsection{Environment}
\subsection{Markov Decision Process of Multi-constrained EVRP}

In this paper, we model the multi-constrained EVRP as a two-player game, then the Markov decision
process~(MDP) can be defined as a tuple $(\mathcal{S},\mathcal{A},\mathcal{R},\mathcal{P}) $, consisting of state $\mathcal{S}$, action $\mathcal{A}$, reward $\mathcal{R}$ and state transition $\mathcal{P}$.
\begin{itemize}
\item State: $\mathcal{S}$ represents the state space. In a two-player game, $\mathcal{S}=(s_t^1,s_t^{-1})$, each player starts from the depot with the initial state $s_0^1$, $s_0^{-1}$, respectively, and $s_t^1$, $s_t^{-1}$ represent the states of two players at time step $t$. The graph node state and the electric vehicle state, represented as $\{ s_t^1$, $s_t^{-1}\} =\{x_{t}^1,v_{t}^1,x_{t}^{-1},v_{t}^{-1}\}$ . For each node $i$, $x_t^i = (x_s^i,c_t^i)$, $x_s^i,c_t^i$ are the static and dynamic information of the node $i$, respectively. The static information is composed of the two-dimensional coordinates of the node $x_s^i$ and the demand of each customer $c_t^i$. For the vehicle state $ \{v_{t}^1,v_{t}^{-1}\}=\{e_t^1,\tau_t^1,u_t^1,e_t^{-1},\tau_t^{-1},u_t^{-1}\}$, $e_t$ is the remaining battery of the electric vehicles; $\tau_t$ is the current travel time; and $u_t$ is the remaining capacity of the electric vehicles.
\item Action: $\mathcal{A}$ represents the action space. In the two-player game $\mathcal{A}=\{a_0^1,a_1^1,\cdots,a_T^1\ ,a_0^{-1},a_1^{-1},\cdots,a_T^{-1}\}$, the action $a_t$ represents the action that has been chosen at the time step $t$.
\item Reward: Unlike a single task that sets the reward as minimization or maximization of the objective function, the reward $\mathcal{R}$ is reshaped into a binary ${\pm1}$ based on self-competition, which we compare the trajectory $\zeta_0^p=(s_0^p,a_0^p,\ldots,s_{T-1}^p,a_{T-1}^p,s_T^p)$ for the player $p\in\{1,-1\}$ at the time step $t$, $\mathcal{R}=0$ if $t<T-1$.
\item Transition: The state transitions deterministically to $\mathcal{P}=F(s_t,a_t)$ due to the deterministic state transition function $F\colon\mathcal{S}\times{\mathcal{A}}\to\mathcal{S}$ .
\end{itemize}
\begin{figure*}[bth!]
\centering
\includegraphics[width=1.0\textwidth]{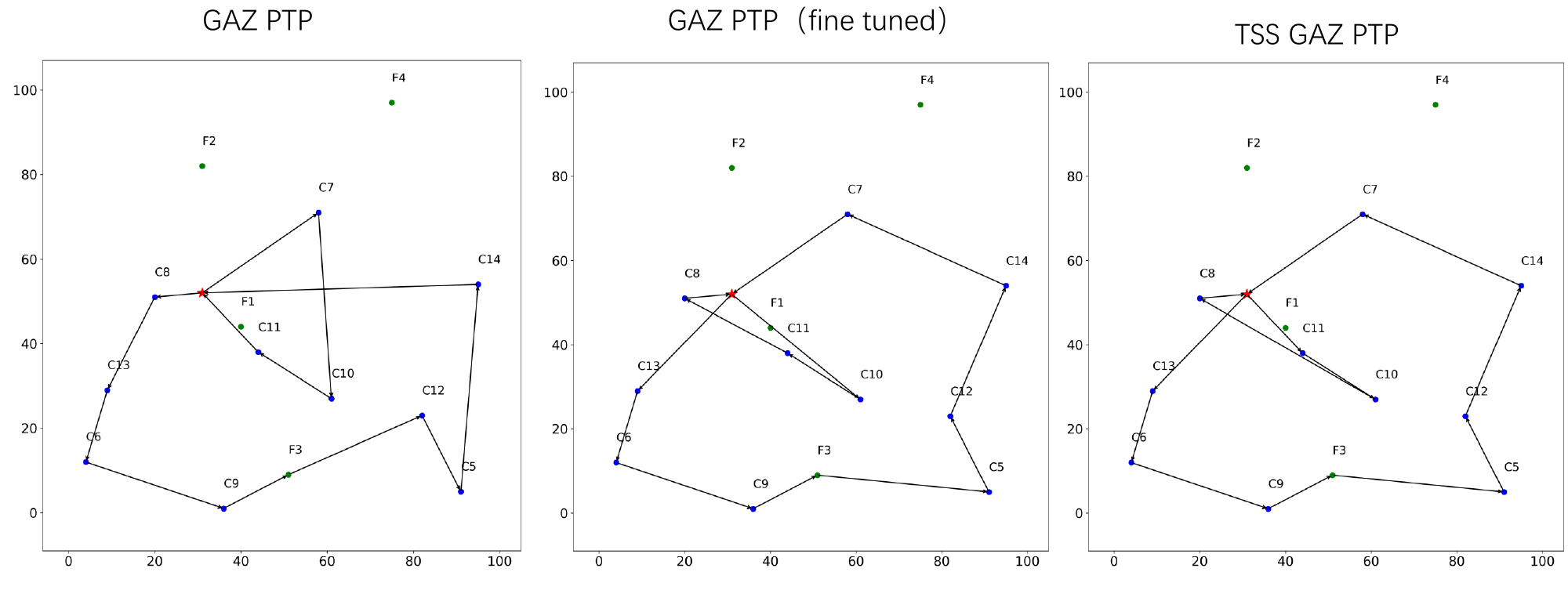} 
\caption{An Example of Optimal Solution Path Visualization of EVRP10 under Different Methods.}
\label{route10}
\end{figure*}
\begin{figure*}[bth!]
\centering
\includegraphics[width=1.0\textwidth]{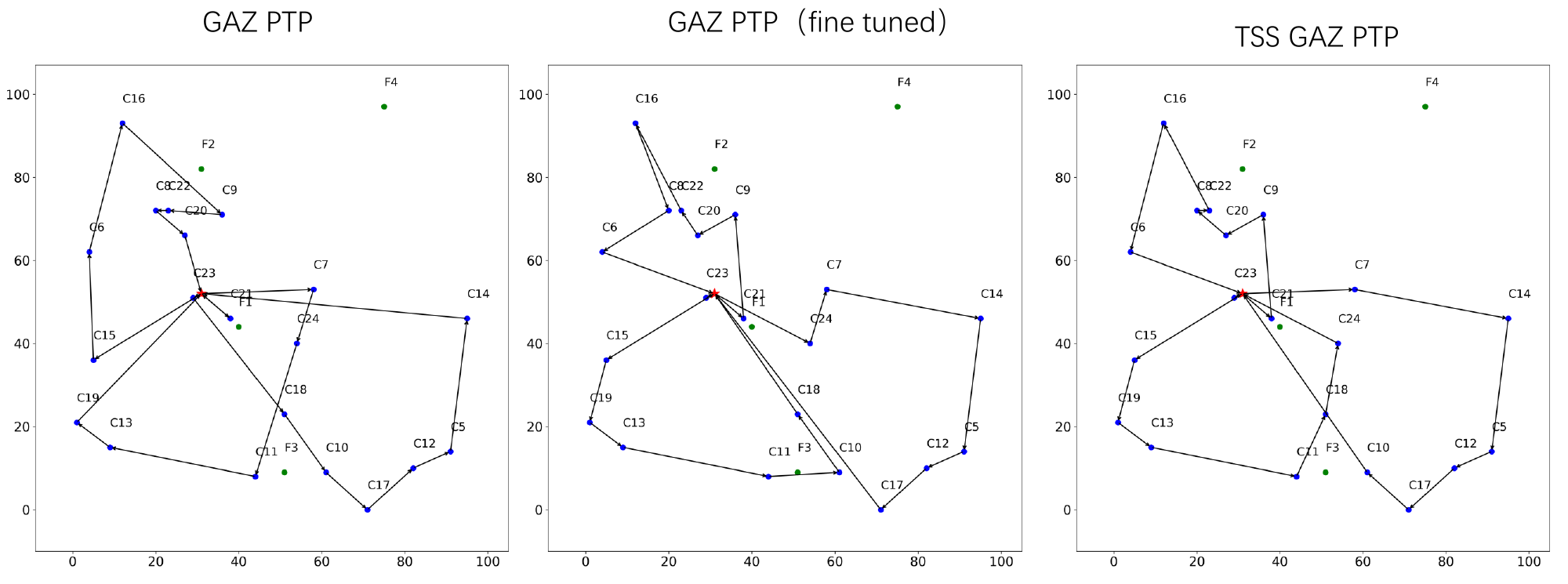} 
\caption{An Example of Optimal Solution Path Visualization of EVRP20 under Different Methods.}
\label{route20}
\end{figure*}
\begin{figure*}[bth!]
\centering
\includegraphics[width=1.0\textwidth]{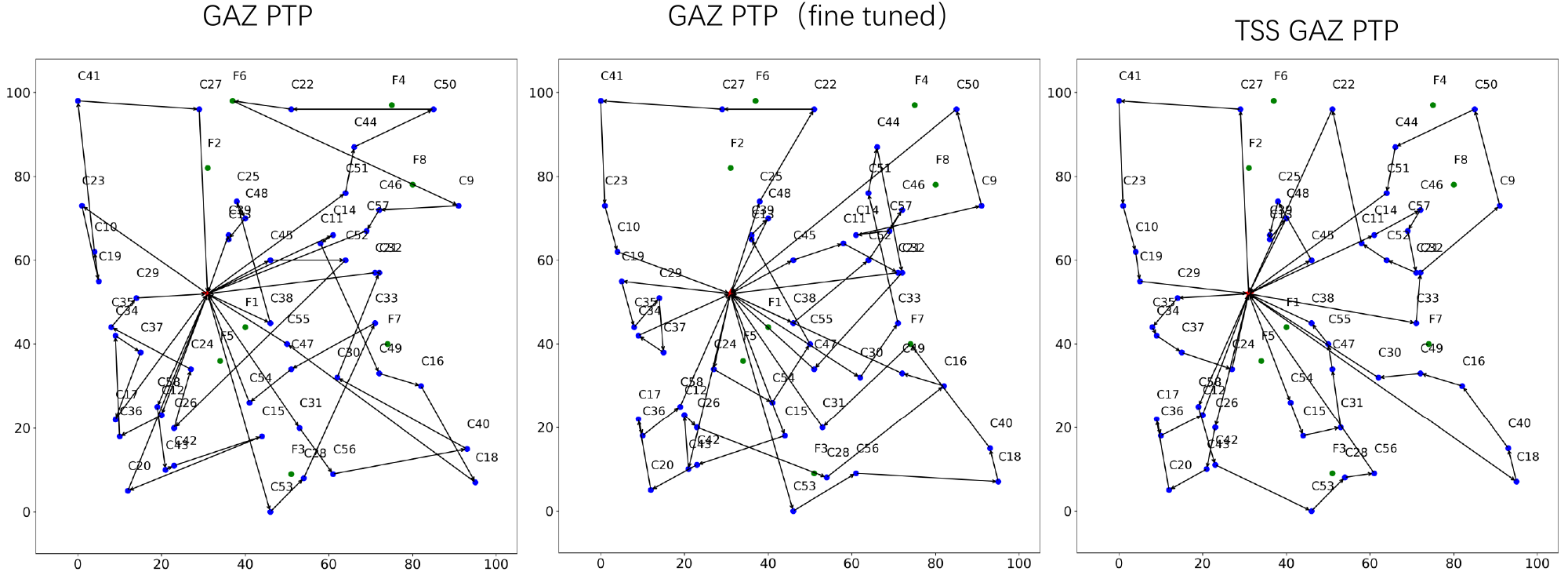} 
\caption{An Example of Optimal Solution Path Visualization of EVRP50 under Different Methods.}
\label{route50}
\end{figure*}

\end{document}